\documentstyle[twocolumn,prb,aps]{revtex}
\begin{document}
\twocolumn[\hsize\textwidth\columnwidth\hsize\csname @twocolumnfalse\endcsname
\draft

\title{Current Carrying States in a
Random Magnetic Field
}

\author{Kun Yang and R. N. Bhatt}
\address{Department of Electrical Engineering, Princeton University,
Princeton, NJ 08544}
\date{\today}

\maketitle

\begin{abstract}
We report results of a numerical study of noninteracting
electrons moving in two dimensions, in the presence of a random potential 
and a random magnetic field for a sequence of finite sizes, using
topological properties of the wave functions
to identify extended states. Our results are consistent with the existence of
a second order localization-delocalization
transition driven by the random potential. The critical randomness strength
and localization length exponent are estimated via a finite size scaling 
analysis.
\end{abstract}
]

The problem of Anderson localization in two dimensions (2D) continues to be the
subject of extensive research recently. Scaling arguments\cite{gangof4}
predict that all one-particle states are localized in 2D, in the presence of
scattering potential which respects time reversal symmetry. 
In the presence of a {\em uniform} magnetic field, however, 
because of the fact that the magnetic field breaks time reversal symmetry
and suppresses the enhanced back scattering,
extended states
are found at individual critical energies.\cite{bodoreview}
These extended states are crucial to the existence of the integer
quantum Hall effect.\cite{bodoreview}

Recently there has been considerable interest in the localization problem in 2D
in the presence of a {\em random} magnetic field with {\em zero} mean, 
motivated by the 
work of Halperin, Lee and Read,\cite{hlr}
and Kalmeyer and Zhang,\cite{kz} on the problem of the half filled
Landau level. These authors argue that if one treats the electrons as 
composite fermions\cite{jain} carrying two flux quanta, the flux
carried the the composite fermions cancels that of the external magnetic
field {\em on average} when the Landau level filling factor is $1/2$, and the
systems may be viewed as a Fermi liquid like metal. 
Disorder induces local fluctuations in the density of the composite fermion,
and hence at the mean
field level the composite fermions see a {\em fluctuating}
local magnetic field with zero mean, in the presence of disorder.

Following this development, there have been many efforts devoted
to the study of the 
localization problem of noninteracting particles moving in a random magnetic
field with zero 
mean.\cite{sn,ahk,kwpz,lc,zhang,amw,liu,sheng,kim,miller,kramer}
However, the key issue, namely whether extended states exist
in the thermodynamic limit, remains 
unresolved. Zhang and Arovas\cite{zhang} argued that 
in the field theory description of the problem,
the local magnetic
flux gives rise to a local topological density which has long range interaction
and induces a 
localization-delocalization transition (see, however, Ref. [\onlinecite{amw}]).
Their conclusion is supported by numerical work of several 
groups\cite{ahk,kwpz,liu,sheng} which found extended states near the band 
center on lattice models with weak random potential and finite random 
magnetic field, although 
Sugiyama and Nagaosa\cite{sn} concluded otherwise. On the other hand, 
Lee and Chalker\cite{lc} simulated a network model which they argue to be
appropriate to the problem of random magnetic field, and find states are 
localized at all energies. Subsequently Kim, Furusaki and Lee\cite{kim} showed
that the network model of Lee and Chalker may be mapped onto an 
$SU(2N)$ spin chain in the limit $N\rightarrow 0$, which has short range
correlations and an excitation gap; they thus confirm the conclusion of 
Lee and Chalker within the framework of the network model. 
The network model was originally
introduced\cite{chalker}
in the study of the integer quantum Hall effect, and has enjoyed
tremendous success in that context. Such success is ensured by the fact that
in the presence of a strong {\em uniform} magnetic field, the magnetic 
length is {\em very small}, thus one body states live along 
{\em one dimensional} equipotential lines. In the case of a {\em random}
magnetic field, however,
the relevant percolating equal field lines are along $B=0$, where the
magnetic length is {\em infinite}! Therefore it is not completely obvious
that the network model of Lee and Chalker is an appropriate description
of the random
magnetic field problem.

In this paper we report a systematic finite size numerical study 
of the electron
localization problem in the presence of a random magnetic field with zero
mean, as well as a random potential, on the square lattice. We study 
topological properties of one-electron wave functions by calculating their
topological quantum numbers called the Chern number.\cite{tknn,niu}
This approach has proved very successful in the study of integer quantum Hall 
effect and transitions.\cite{arovas,huo,yang}
In particular, finite size corrections appear to be much less in this
approach, compared with others, implying a rapid convergence to the
thermodynamic limit. Recently it has been applied by Sheng and
Weng\cite{sheng} to the present problem. There are, however, several 
important differences between our work and theirs. Firstly, their work 
as well as other
numerical studies,\cite{sn,ahk,kwpz,liu} 
concentrate on systems with a relatively {\em weak} (or zero)
random potential, and attempt to identify a mobility edge. When the
strength of random potential is weak, the length scales involved may be 
quite large,
and it is often hard to distinguish whether the localization length
near the band center
is truly infinite or very large but finite. In contrast,
we study systems with a series of different random potential strengths,
with the same strength of random {\em flux}. The idea is that {\em if} 
extended states do exist for weak randomness, there must exist a 
critical random potential strength (which is typically
{\em not} very weak) at which all extended states disappear,
and scaling behavior should be observable near this critical point.
Secondly, unlike Sheng and Weng, who study 
behavior of density of states ($\rho(E)$) and
density of current carrying
states ($\rho_c(E)$), which are functions of energy $E$,
we focus on quantities like
the {\em total number} of extended states $N_c$
in a finite size system. $N_c$ is the
zeroth moment of the distribution function $\rho_c(E)$. 
In other random systems such as spin glasses low order moments of distribution
functions are
known to converge to an acceptable level of accuracy 
relatively fast with sample averaging; the higher 
moments and the distributions themselves are still quite noisy at that stage.
Finally, our massively parallel computer allows us to study samples of
considerably larger sizes than those in Ref. [\onlinecite{sheng}].

We consider the one body tight binding Hamiltonian on the square lattice:
\begin{equation}
H=-\sum_{\langle ij\rangle}(e^{ia_{ij}}c_i^{\dagger}c_j+
e^{-ia_{ij}}c_j^{\dagger}c_i)+\sum_i\epsilon_ic_i^{\dagger}c_i,
\end{equation}
where $c_i$ is the fermion operator on lattice site $i$. The first term
represents hopping or kinetic energy of the electrons, and the summation 
is over nearest neighbors. The flux through each square 
is equal to the summation of $a_{ij}$ along its four edges. We take 
the flux to be
random and uniformly distributed from $-h\pi$ and $h\pi$, where $0\leq h\leq 1$.
The second term represents a random onsite potential. We take $\epsilon_i$
to be uniformly distributed between $-w$ and $w$. 
$h$ and $w$ parameterize the strength of random magnetic field and
random potential respectively; when $h=1$ the flux through each square is
completely random. We have set the hopping
matrix element to be 1 and use it as the unit of energy.
We study samples of square geometry with linear size $L$
(with number of sites $N_s=L^2$), and impose 
periodic boundary conditions in both directions:
$\Psi(k+L\hat{x})=e^{i\phi_1}\Psi(k)$, and 
$\Psi(k+L\hat{y})=e^{i\phi_2}\Psi(k)$.
The Hall conductance of an individual eigenstate $|m\rangle$ can be
obtained easily using the Kubo formula:
\begin{equation}
\sigma_{xy}^{m}={ie^2\hbar\over N_s}\sum_{n\ne m}{\langle m|v_y|n\rangle
\langle n|v_x|m\rangle-\langle m|v_x|n\rangle\langle n|v_y|m\rangle\over
(E_n-E_m)^2},
\end{equation}
where $E_n$ is the energy of the $n$th state, and 
\begin{equation}
v_{\tau}=(i/\hbar)\sum_j(c^{\dagger}_{j+\tau}c_je^{ia_{j+\tau,j}}
-c^{\dagger}_{j}c_{j+\tau}e^{-ia_{j+\tau,j}})
\end{equation}
is the velocity operator along direction $\tau$ ($=\hat{x}$ or $\hat{y}$).
$\sigma_{xy}$ is identically zero in the absence of magnetic flux,
as guaranteed by time reversal symmetry. 
In the presence of random magnetic flux with zero average, the time reversal
symmetry is broken in a specific configuration of randomness, and
{\em individual} states may have nonzero $\sigma_{xy}$,
although the {\em disorder averaged} Hall conductance is always zero for any
given Fermi energy, since the averaging process restores the time reversal
symmetry. $\sigma_{xy}^m$ depends on the
boundary conditions imposed. As shown by Niu {\em et al.},
the boundary condition averaged Hall conductance takes the
form\cite{niu}
\begin{equation}
\langle\sigma_{xy}^m\rangle={1\over 4\pi^2}\int{d\phi_1d\phi_2\sigma_{xy}^{m}
(\phi_1,\phi_2)}=C(m)e^2/h,
\label{chern}
\end{equation}
where $C(m)$ is an integer called the Chern number of the state $|m\rangle$.
States with nonzero Chern numbers carry Hall
current and are necessarily
extended states.\cite{arovas} Thus by numerically diagonalizing
the Hamiltonian on a grid of $\phi_1$ and $\phi_2$, and calculating the
Chern numbers of states of finite size systems by converting the integral in
(2) to a sum over grid points, we are able to identify
extended states in a {\em finite size} system unambiguously.

In this work we have studied finite size samples with linear size ranging
from $L=4$ to $L=19$, for different random potential and random field
strength. For each particular randomness strength and system size, we 
average over
30 to 2000 different configurations of randomness. Depending on system size
and randomness strength, the number of grids 
necessary to determine the Chern numbers unambiguously
using Eq. (\ref{chern})
ranges from $35\times 35$ to $70\times 70$. 

We plot in Fig. 1 the dependence of disorder averaged
number of current carrying states
$N_c$ versus the system size $N_s$, for $h=0.5$ and a series of different
$w$, on a {\em double logarithmic} plot. We find for a given system size, 
$N_c$ decreases as $w$ increases. This is exactly what is expected, 
since random potential tends to localize states. 
For very strong randomness ($w\ge 4.5$)
and big enough system size, we see $N_c$
decreases with increasing $N_s$, indicating there are no extended states
in the thermodynamic limit, and the localization length is finite throughout
the band. For somewhat weaker randomness, {\em e.g.} $w=4.0$, although 
$N_c$ increases with $N_s$ within the size range we are able to reach,
the curves obviously have downward curvature, and  
beyond certain $N_s$, $N_c$ presumably decreases with increasing size
and eventually goes 
to zero. Current
carrying states appear in {\em finite size} systems in this case
because the localization length $\xi$ for states near the band
center, although finite, may be larger than the system size
$L$.\cite{huo}
For even weaker randomness ($w<3.0$), however, we do not see any 
evidence of such downward curvature, and it appears that $N_c\rightarrow\infty$
as $N_s\rightarrow\infty$.
Such behavior is expected if extended states exist in the thermodynamic limit,
either at individual critical energies,\cite{huo,yang} or in a finite width
band near the band center.
In both cases, one expects the asymptotic behavior on the double logarithmic
plot tp be linear. In the former case, the slope is expected to be less than
$1$, while for a band of extended states we have $N_c\propto N_s$,
hence the slope equals unity.
For weak enough
random potential, $w\le 2.0$, we do find that the slope $y$ is extremely
close to $1$, consistent with 
a band of extended states
in the thermodynamic limit. 

It has been suggested\cite{amw} that for weak randomness states are just 
barely localized (as in the case without magnetic field\cite{gangof4})
with exponentially large $\xi$ near the band center, much larger than
system sizes accessible to numerical studies. While this may give the
impression of an extended band, one would expect a slow, logarithmic 
decrease in $N_c/N_s$ with increasing system size. However, we find no
such evidence 
for small $w$; $N_c/N_s$ appears to be a constant, or
{\em increases} slightly.
If, on the other hand, there is a band of extended states for small $w$, the
width of such a band should decrease as $w$ increases and shrink to zero
at a critical random potential $w_c$, above which $\xi$ is finite throughout
the band (see Fig. 2). For $w<w_c$ we should have $N_c\propto N_s$ as 
$N_s\rightarrow\infty$,
while for $w>w_c$, $N_c\rightarrow 0$ in this limit.
Right at the critical point $w=w_c$, there is a {\rm single} critical 
energy at which the localization length diverges, just like the case of a
uniform magnetic field,\cite{huo,yang}
and we should have $N_c\sim N_s^{y_c}$,
where $y_c$ is an exponent related to the localization length
exponent $\nu$ through\cite{huo,yang,note3}
\begin{equation}
y_c=1-{1\over 4\nu}.
\label{y}
\end{equation}
For $w$ close to $w_c$, there is a characteristic length scale that
scales as $\xi_m\sim |w-w_c|^{-\nu}$. 
For $w>w_c$, $\xi_m$ is the maximum localization length of the system, 
while for $w<w_c$, $\xi_m$ is the length scale above which 
the behavior $N_c\propto N_s$ is seen. Near the critical point we should see
scaling behavior: 
\begin{equation}
N_c\sim N_s^{y_c}\tilde{F}^{\pm}(L/\xi_m)\sim N_s^{y_c}F^{\pm}(N_s
|w-w_c|^{2\nu}),
\label{scale}
\end{equation}
where $F^{\pm}(x)$ are two different scaling functions for
$w<w_c$ and $w>w_c$ respectively, which should satisfy the asymptotic
behavior $F^+(x)\sim x^{1-y_c}$ and $F^-(x)\rightarrow 0$ as 
$x\rightarrow\infty$.
Therefore, {\em if} there is indeed a critical randomness $w_c$, we 
ought to see
scaling behavior (\ref{scale}), by tuning a {\em single} parameter 
$w_c$. The other parameter $\nu$ in the scaling relation (\ref{scale}) may be
determined using Eq. (\ref{y}), since the slope $y$ is know from the data
presented in Fig. 1.

Such scaling behavior is indeed seen, and the data collapses best 
for $w_c=3.0$ and $\nu=1.8$, as shown in Fig. 3. 
At $w=3.0$ the linear fit of
$\log N_c$ versus $\log N_s$ yields $y=0.87\pm 0.01$, and from Eq. (\ref{y})
we obtain $\nu=1.9\pm 0.2$, consistent with scaling results. We emphasize
that data collapsing of {\em two} different scaling curves is achieved 
by tuning {\em one} parameter, due to the constraint Eq. (\ref{y}). 
The scaling behavior we see here supports
the existence of a band of extended states for weak random potential,
and would not be expected if there were no localization-delocalization 
transition at finite $w_c$.
In order to make sure that our data are indeed in the scaling regime,
we plot in Fig. 4 both the total density of states $\rho(E)$ and density of
current carrying states $\rho_c(E)$ (per site), 
for $w=3.0$ and $h=0.5$, for two
different system sizes $L=4$ and $L=16$. We find 
the width of $\rho_c(E)$
is considerably smaller than that of $\rho(E)$,
especially for the larger size. Further the width of $\rho_c(E)$ is
size dependent while that of $\rho(E)$ is essentially the same for
both sizes. This is similar to what is
seen in the study of the localization problem in lowest Landau level,\cite{huo}
and indicates that we are indeed in the scaling regime.
By contrast the width of $\rho_c(E)$ and $\rho(E)$ is almost the same
for Gaussian white noise potential in the first Landau level, 
where the localization lengths
are known to be very large.\cite{bodoreview,guo}

From the scaling we are also able to estimate the critical randomness to
be $w_c=3.0\pm 0.2$, and $\nu=1.8\pm 0.2$, for $h=0.5$.
The localization length exponent $\nu$ we obtain here is very close to that
estimated by Sheng and Weng\cite{sheng} using a very different scaling
scheme, although they do not have a quantitative estimate of the critical
random potential strength $w_c$.

In summary, we have found good evidence supporting the existence
of a localization-delocalization transition driven by random potential, in the
2D square lattice in the
presence of a random magnetic field with zero average, implying the
existence of extended states at weak randomness. The critical 
random potential strength and localization length exponent are estimated
using finite size scaling analysis.

We thank Yong Baek Kim, Z. Y. Weng, and P. W\"olfle
for helpful discussions. This work
was supported by NSF grants DMR-9400262 and CDA-9121709. R.N.B. was also
supported in part by a Guggenheim fellowship, and thanks the Aspen Center
for Physics for hospitality while this manuscript was being written up.

\begin{figure}
\caption{Number of extended states $N_c$ versus system size $N_s$, for $h=0.5$
and
various $w$ on a double logarithmic scale.}
\end{figure}

\begin{figure}
\caption{Schematic phase diagram of the random flux problem}
\end{figure}

\begin{figure}
\caption{The scaling functions $F^{\pm}(N_s|w-w_c|^{2\nu})$.}
\end{figure}

\begin{figure}
\caption{Disorder averaged density of states $\rho(E)$ and density of
current carrying states $\rho_c(E)$ at $w=3.0$ and $h=0.5$, for $L=4$
and $L=16$.}
\end{figure}

\begin{references}
\bibitem{note3} This relation is slightly different from the uniform field
case,\cite{huo,yang} in which case one has $y=1-1/(2\nu)$. This is
because of the fact in that case there are only individual critical energies
near which $\xi\sim |E-E_c|^{-\nu}$, for $w<w_c$.\cite{yang}
In the present case however, at $w=w_c$,
the distance to criticality is proportional to $E^2$, instead of
$E$ ($E_c=0$),
due to the fact that there is a {\em band} of extended states for $w<w_c$.
Thus $\xi\sim |E-E_c|^{-2\nu}$ for $w=w_c$. For $w<w_c$ we still have
$\xi\sim |E-E_c|^{-\nu}$, where $E_c$ is the mobility edge.
See Fig. 2.
\end{references}
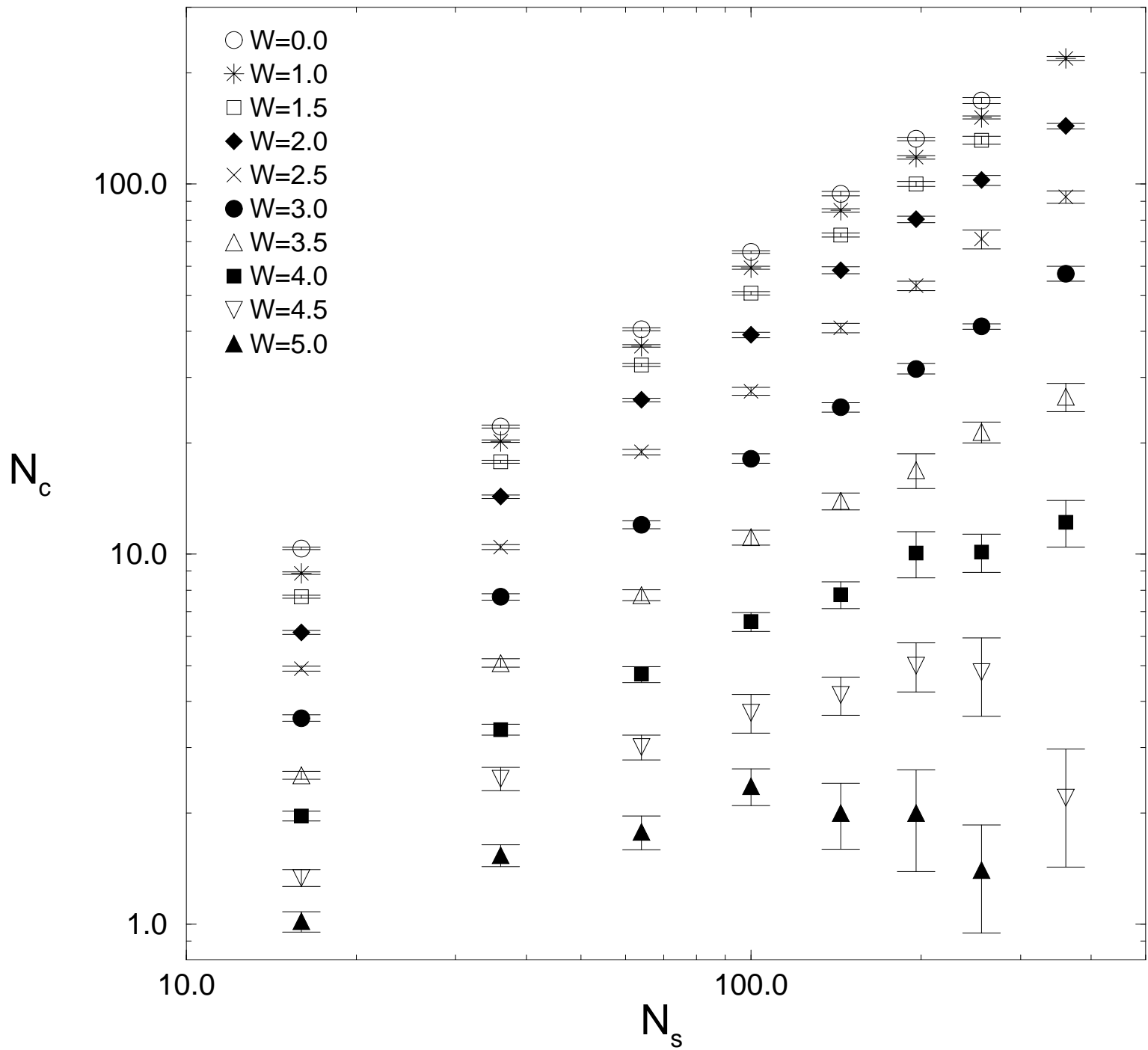
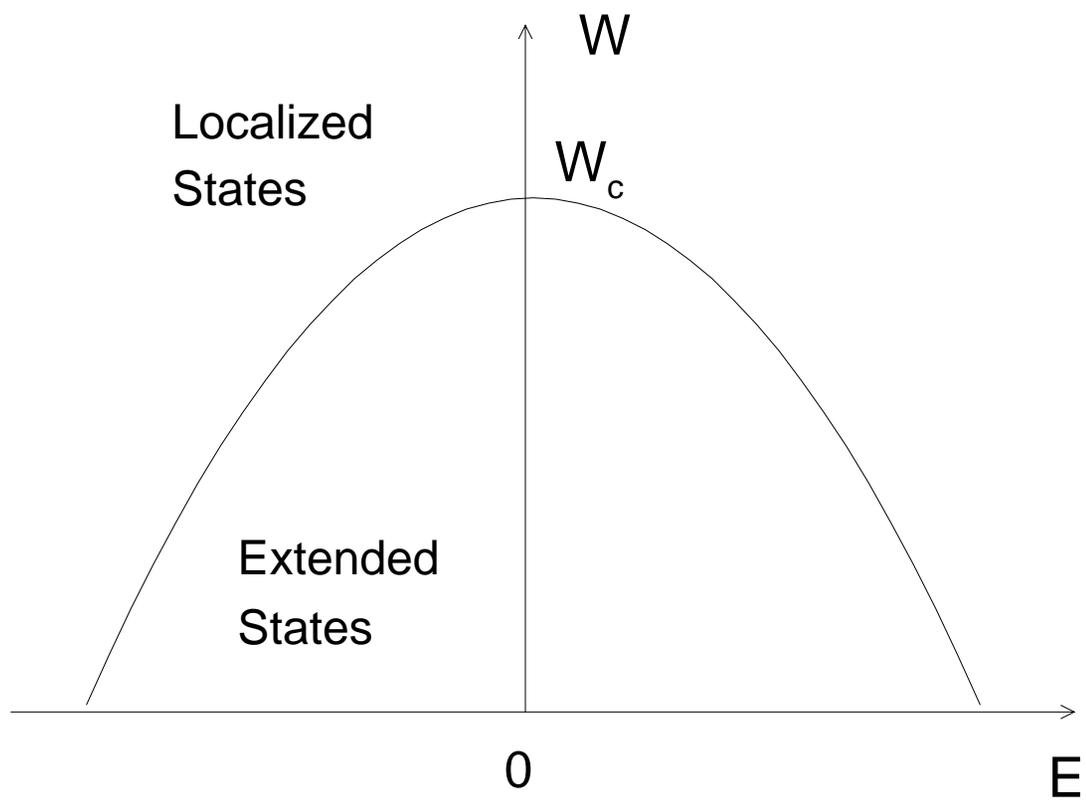
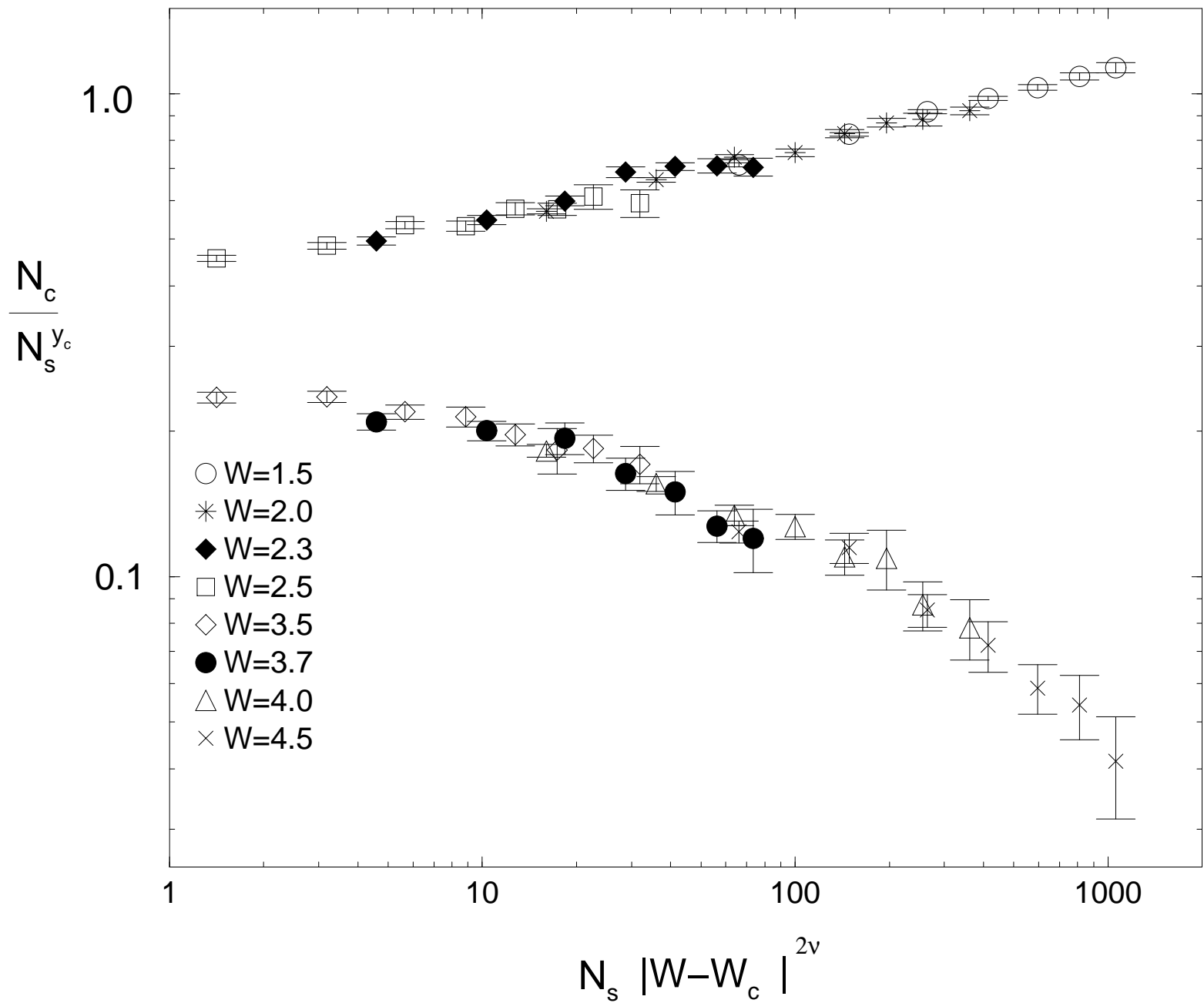
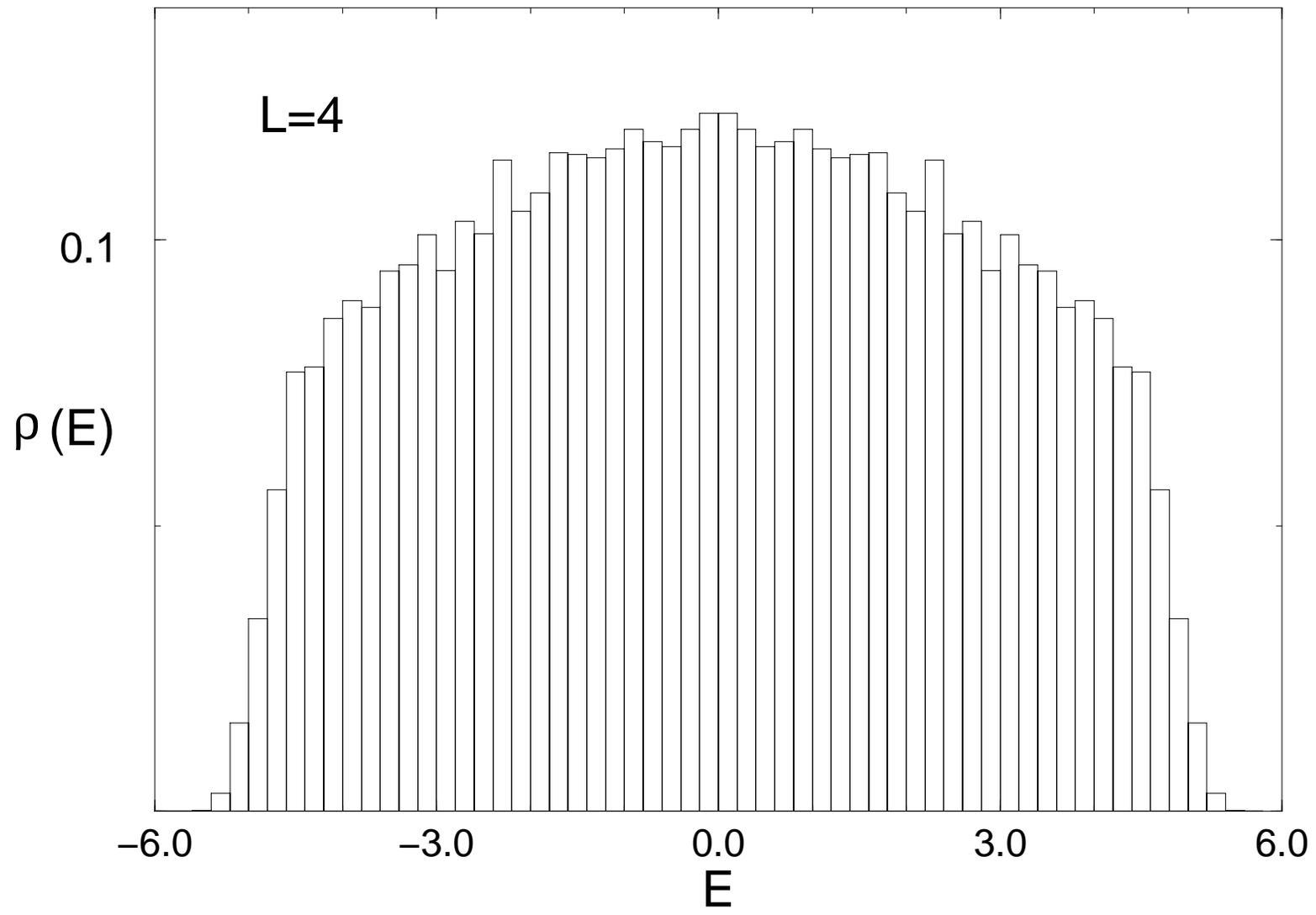
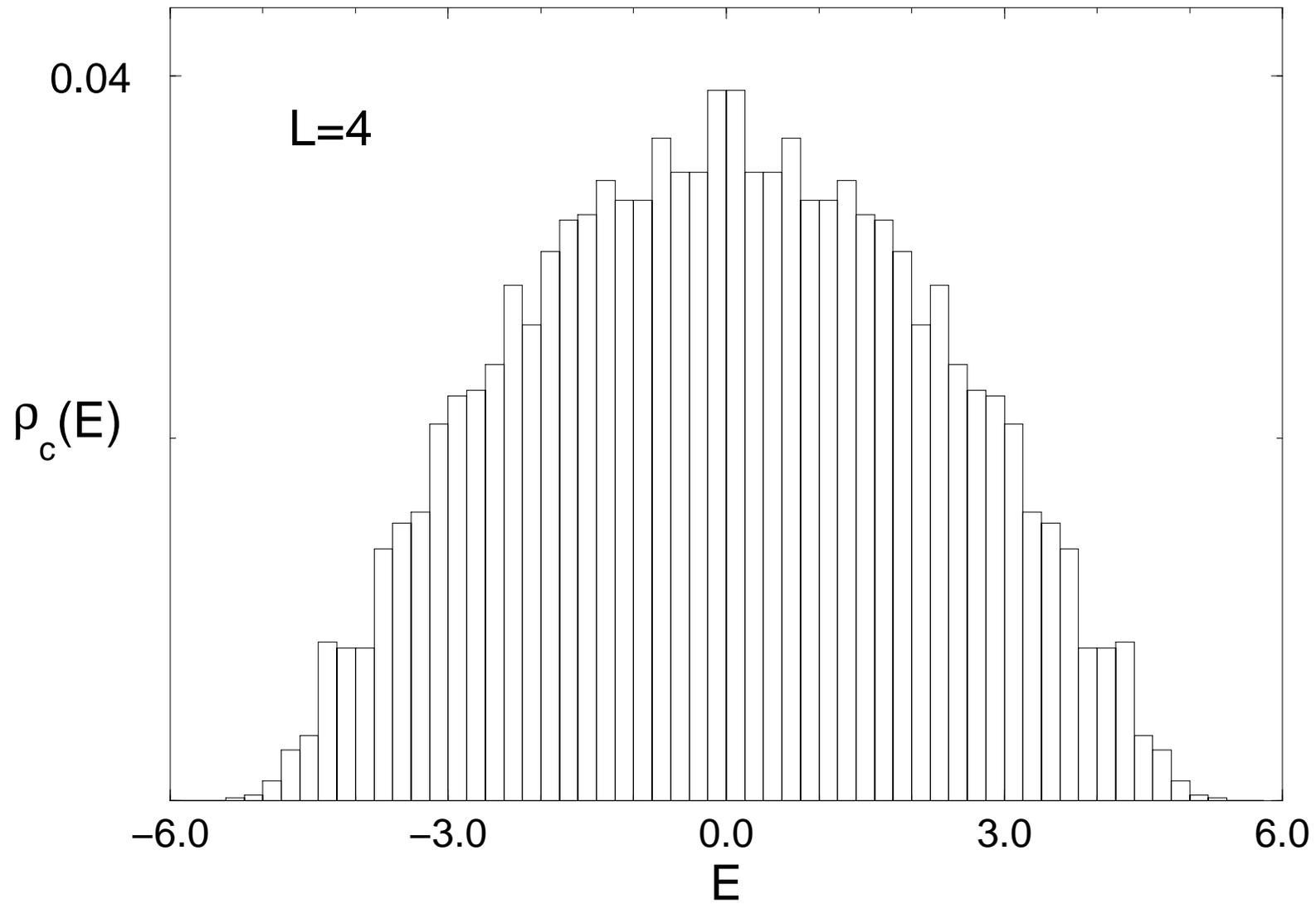
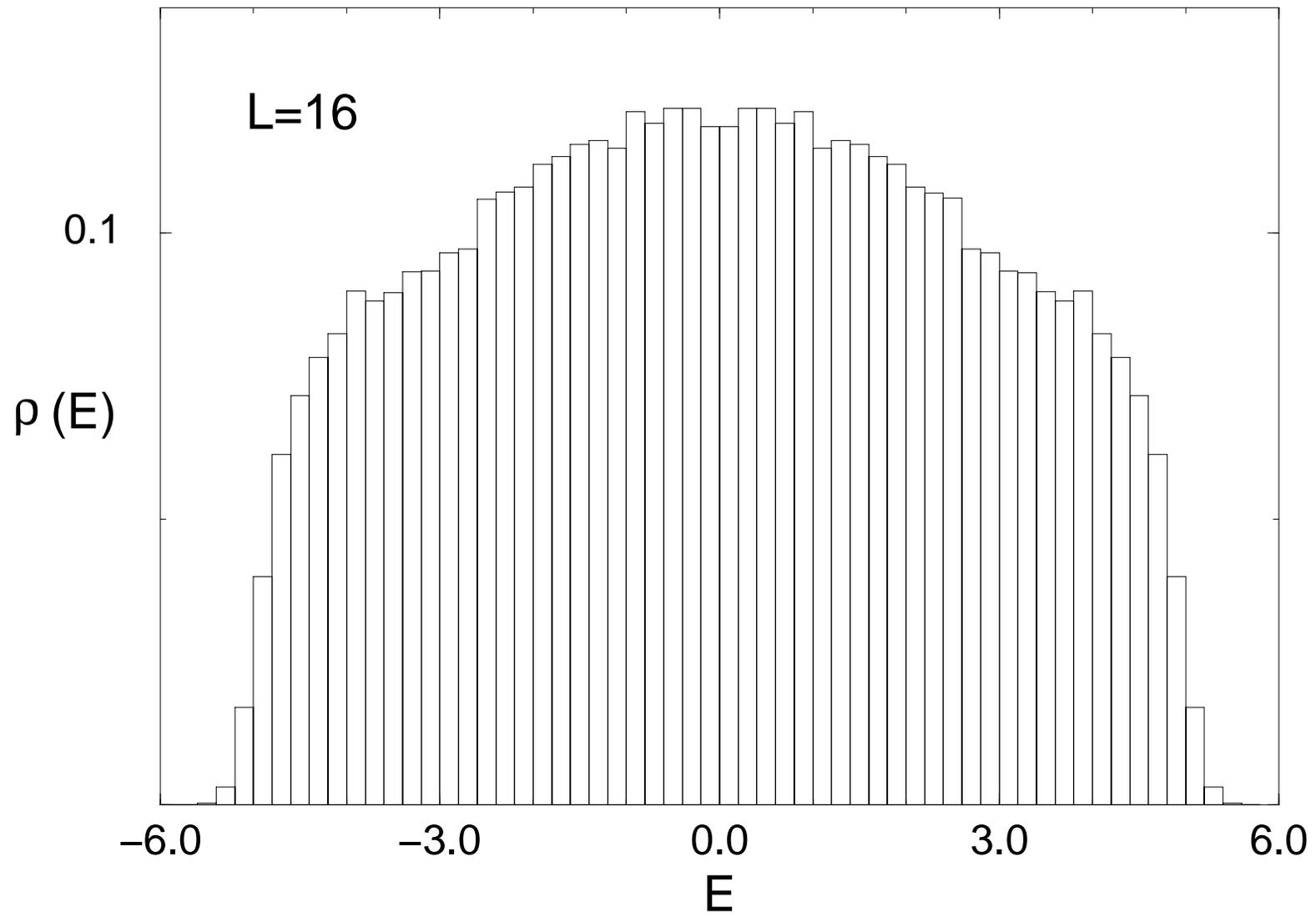
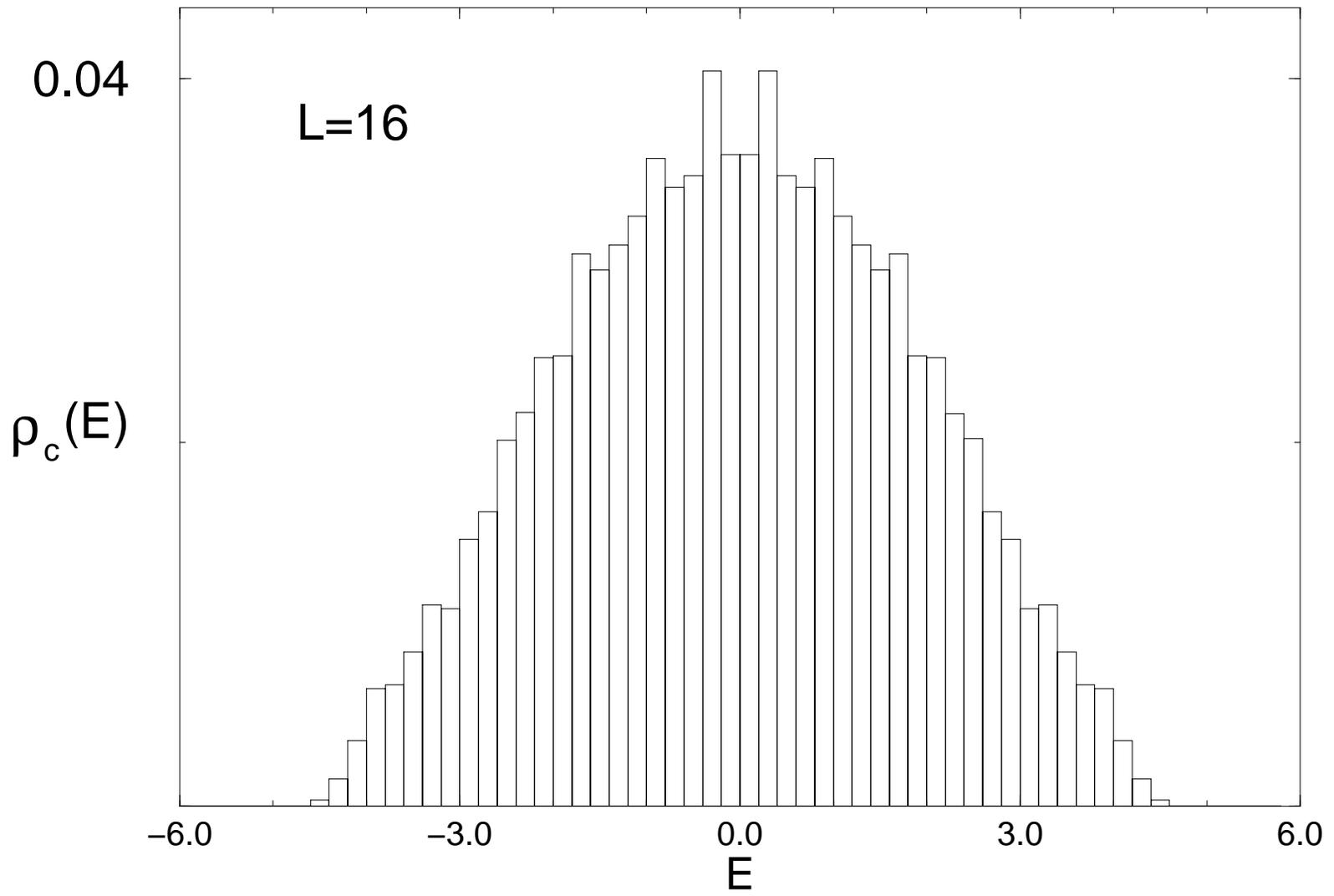
\end{document}